\documentclass[11pt,twoside]{article}

\usepackage{asp2014}

\aspSuppressVolSlug
\resetcounters

\begin{document}

\title{FAIR approach for Low Frequency Radio Astronomy}

\author{Baptiste Cecconi,$^1$}
\affil{$^1$LESIA, Observatoire de Paris-PSL, CNRS, Sorbonne Univ., Univ. Paris Cit\'e, Meudon, France; \email{baptiste.cecconi@obspm.fr}}

\paperauthor{Baptiste 
Cecconi}{baptiste.cecconi@obspm.fr}{0000-0001-7915-5571}{Observatoire 
de Paris-PSL}{LESIA}{Meudon}{}{92190}{France}



\begin{abstract}
The Open Science paradigm and the FAIR principles (Findable, Accessible, Interoperable, 
Reusable) are aiming at fostering scientific return, and reinforcing the trust in science 
production. The MASER (Measuring, Analysing and Simulating Emissions in the Radio range) 
services implement Open Science through a series of existing solutions that have been put 
together, only adding new pieces where needed. It is a ``science ready'' 
toolbox dedicated to time-domain low frequency radioastronomy, which data products mostly 
covers solar and planetary observations. 

MASER solutions are based on IVOA protocols for data discovery, on IHDEA tools for data 
exploration, and on a dedicated format developed by MASER for the temporal-spectral 
annotations. The service also proposes a data repository for sharing data collections, 
catalogues and associated documentation, as well as supplementary materials associated to 
papers. Each collection is managed through a Data Management Plan, which purpose is two-fold: 
supporting the provider for managing the collection content; and supporting the data centre for 
resource management. Each product of the repository is citable with a DOI, and the landing page 
contains web semantics annotations (using schema.org).\end{abstract}



\section{Low Frequency Radio Astronomy}
The planets of the solar system are obstacles in the flow of the extended atmosphere of the 
Sun, the solar wind. The amplitude of this interaction depends on the size of the obstacle, 
hence interactions with magnetized planets result in much more energetic phenomena 
\citep{zarka_EAS_10}. The planetary magnetospheres are the region in space surround a 
magnetized planet and where the dynamics of the medium is controlled by the planetary magnetic 
field and other internal sources of plasma. Magnetospheres are the largest ones of the solar 
system: the tail of the Jovian magnetosphere can go as far as the orbit of Saturn 
\citep{Desch:2007bx}. They are also accelerating particles through various plasma instabilities 
and dynamical processes, which in turn can produce intense low frequency radio emissions 
\citep{zarka_ASR_92}. The solar activity can also drive low frequency radio emissions, mostly 
known as Solar type II and type III bursts, which are respectively related to shocks 
propagating in the interplanetary medium, and relativistic electron beams. 
Low frequency radio emission are ranging from a few kHz to a few tens of MHz. The 
terrestrial ionospheric cut-off frequency (at $\approx$10 MHz) is blocking any radio wave
below this threshold, requiring space observations for the lower frequencies. 

Low frequency radio emissions are resulting from plasma instabilities, transferring the free 
energy available in unstable plasma (such as electron beams) into electromagnetic 
radiation. These are non-thermal radiation processes (i.e., not related to atomic and molecular 
transitions), very intense 
and highly sporadic. The time scales of such emissions range from milliseconds (duration of 
individual fine scale bursts, like, e.g., the radiation produced by single electron beams), to 
minutes (duration of an event), with longer scales related to the planetary rotation periods 
(several hours), solar wind cycle fluctuations (month) or seasonal effects (years). 

Since part of the low frequency radio spectrum is not observable from ground, space 
instrumentation have been developed since the beginning of space exploration 
\citep{alexander_AA_75}. Classical radio astronomy space instrumentation is based on one 
\citep[see, e.g.,][]{gurnett_SSR_92,Kurth:2017fv} to three electric 
\citep[see, e.g.,][]{bougeret_SSR_95,gurnett_SSR_04} dipoles on spinning or stabilized 
platforms, allowing to reconstruct the incoming radio waves properties from single point 
observations \citep{Cecconi:2011wo}. These techniques are know as ``Direction Finding'' or 
``goniopolarimetry'', and they provide scientists with the direction of arrival of the radio 
wave, its flux density and its polarization state. Ground based observatories are usually more 
sensitive than space instrumentation, thanks to their larger (and multiple) antenna, using beam-
forming or interferometry techniques, such as, e.g., LOFAR \citep{vanHaarlem:2013gi} or NenuFAR 
\citep{Zarka:2020tc}. 

Recently, similar radio emissions phenomenology have been identified 
while observing an active star \citep{Zhang_2023}, or in stellar systems hosting 
exoplanets \citep{Turner:2021bf}, opening up a new era of plasma instability remote sensing in 
stellar and planetary environments. 

As a summary, low frequency radio astronomy are tracers of unstable energetic charged 
particles (e.g., electron beams, out of equilibrium particle distributions functions), 
most often in magnetised plasma (not always). They are not tracing atomic or molecular lines. 
The radio sources can be electrostatic discharges, plasma waves (with mode conversion), 
cyclotron emission, or synchrotron emission. They are non-thermal emission, usually 
significantly (or fully) polarized, and they are observed with a wide range of time scales: 
milliseconds to years. They are thus a powerful remote sensing tool for observing unstable and 
energetic charged particle populations.

\section{The MASER service as a FAIR-enabling toolbox}

The MASER (Measuring, Analysing and Simulating Emissions in the Radio range, \url{https://maser.lesia.obspm.fr}) service is a 
science ready toolbox  developed at Observatoire de Paris, in France \citep{Cecconi:2020bc}. It 
is recognized as a ``National Observation Service'' by INSU (Institut National des Sciences 
de L'Univers) and is supported by PADC\footnote{Re3data identifier: 
\url{http://doi.org/10.17616/R31NJMS9}} (Paris Astronomical Data Centre). 

MASER data product types are: (i) \emph{spectrograms}, also referred to as \emph{dynamic 
spectra}: they are two-dimensional data sets containing radio wave parameters, such as the flux 
density or the polarization degrees, with dependencies on temporal and spectral domains; (ii) \
\emph{waveforms}: they are high resolution time series, containing direct sampling of electric 
signal temporal fluctuations; and (iii) \emph{events} or \emph{features}: they are time-tagged 
samples or pieces of information, possibly with a spectral or temporal-spectral coverage, often
associated with observational or derived parameters or categorized with name. A waveform 
snapshot is often considered as an event. Note that \emph{imaging} data products are not in the 
scope of MASER.

During the inception of the MASER service, a survey of users' needs have been conducted 
informally, and lead to the following feature: (a) discovery of datasets; (b) online access for 
(pre)visualisation; (c) python library for programmatic access; (d) annotation and  sharing of 
event or feature catalogues. In a later stage, two other needs have been added: (e) running 
models for interpreting or planning observations; and (f) hosting datasets (for, e.g., supplementary materials).

The science topics of MASER are at the crossroads of several communities, which all have their 
own standards serving different purposes: \emph{IVOA}\footnote{IVOA: \url{http://ivoa.net}} 
(International Virtual Observatory Alliance), which is interoperability driven (schemas, protocols, 
vocabularies); \emph{IPDA}\footnote{IPDA: \url{https://ipda.jpl.nasa.gov}} (International 
Planetary Data Alliance), which is archive driven (information model based on OAIS);  
\emph{IHDEA}\footnote{IHDEA: \url{https://ihdea.net}} (International Heliophysics 
Data Environment Alliance), which is reuse driven (data/metadata formats, protocols, tools);
\emph{OGC}\footnote{OGC: \url{https://www.ogc.org}} (Open Geospatial Consortium), which is 
information modeling and reuse driven (mostly for Earth and planetary surfaces);
\emph{Datacite}\footnote{Datacite: \url{https://datacite.org}}, which is reference driven 
(reference, citation, related resources (DOI provider). A subset of protocols and standards 
from these communities have been selected for MASER, and 
used on most of the data served by MASER. 

\subsection{Data discoverability (IVOA solution)}
Data discoverability is implemented using \emph{EPN-TAP} \citep{2022ivoa.spec.0822E}, an IVOA 
recommendation dedicated to data discovery of solar system science products. EPN-TAP provides 
the science community with a uniform search interface over many data providers. A typical query 
for MASER related product would be: products with observation target set to Jupiter and a 
spectral range below 50 MHz. Thanks to EPN-TAP clients, such as the VESPA (Virtual European 
Solar and Planetary Access) portal \citep{2020DSJ....19...22E}, it is possible to senf the same 
query to several services at once and collect results in the same interface. The MASER EPN-TAP 
services also implement \emph{datalink} \citep{2015ivoa.spec.0617D}, which allows to link data 
products to associated services, quicklooks and documentation. 

\subsection{Remote Access to data (IHDEA solution)}
Some MASER data sets have large data rate (several TBs of data per day) or are covering
long times intervals (up to 20 years, with a few second temporal resolution). There is thus
a need for an optimized distribution system. The \emph{das2} system \citep{2019esoar.10500359P}, 
developed by the University of Iowa (USA) have been implemented on the data collections. They 
can be accessed using Autoplot \citep{2010arXiv1004.2447F}, or with the das2py\footnote{das2py 
Python module: \url{https://github.com/das-developers/das2py}} client. This system was built 
for space data (low data rate), but is capable of serving long resampled times series. It is 
used with success on ground-based Nan\c ay data sets. It is thus also well fitted for ground 
based high data rate data sets. This system implements a run-on-demand server-side resampling 
scheme which optimises data transfers. 

\subsection{Running code on-demand (IVOA solution)}
The UWS protocol \citep{2010ivoa.spec.1010H}, an IVOA recommendation dedicated to 
run-on-demand, has been implemented by the MASER team, so the access to the ExPRES code 
\citep{2019A&A...627A..30L} can be granted. This implementation uses OPUS (Observatoire de 
Paris UWS Server) \citep{2022ASPC..532..451S}. It is currently used by the JUICE (Jupiter Icy 
Moons Explorer) mission ground segment at ESA (European Space Agency) to prepare the science 
segmentation of the orbits at Jupiter \citep{2021P&SS..20905344C}.

\subsection{Python interface (IHDEA solution)}
The \texttt{maser.data} python package is providing a unified access to MASER data sets and 
more (e.g., Cassini/RPWS, Wind/Waves, STEREO/Waves, Juno/Waves, Voyager/PRA, 
Mars-Express/MARSIS, CDPP datasets). It is relying on maintained and acknowledged external libraries 
such as \emph{spacepy} \citep{spacepy11} for managing CDF, \emph{astropy} \citep{astropy:2022}
for managing FITS, and \emph{xarray} \citep{hoyer2017xarray}. The code is available online 
(\url{https://gitlab.obspm.fr/maser/maser4py}) as an open source software. The library is on 
following the PyHC rules \citep{annex_2018_2529131}. It will be soon submitted as a new package 
to PyHC (Python Heliophysics Community). 

\subsection{Annotation and sharing of events (OGC solution)}
There was no prior solution for annotating two-dimensional shapes in the temporal-spectral 
domain. Describing events with a time range and a spectral range was possible, but complex 
shapes (polygons) were not possible to implement. The MASER team has developed a new format, 
based on GeoJSON \citep{butler_geojson}, a standard from the OGC community: the Time-Frequency 
catalogue (TFCat) \citep{10.3389/fspas.2022.1049677}. 

\subsection{Hosting datasets (Datacite solution)}
At the time of writing, the MASER service is hosting about sixty digital records\footnote{See: 
\url{https://maser.lesia.obspm.fr/publications/doi}}, including 
datasets (per instrument, per processing level), associated documentation, TFCat catalogues, 
paper supplementary material, etc. Each published data object has a landing page and a DOI 
(digital object identifier). Datasets are managed with a Data Management Plan (DMP), which is a 
tool to support the providers teams so that they easily: (a) define the structure of their 
collections; (b) select the relevant interoperable interfaces; (c) select standard data formats 
(e.g., FITS or CDF). This a way to empower teams. This helps them to better plan the 
development of the data products, it also helps the hosting data center (PADC) to forecast 
storage management. The MASER team used the DMP template developed by Observatoire de Paris 
\citep{dmp_guide_obspm, dmp_template_obspm}. Such a document is now required by many funders.
The landing pages also include \emph{schema.org} metadata using JSON-LD, which make them 
harvestable by search engines. An open licence is also set so that the product are openly accessible and reusable.  

The DMP documents developed with the MASER team contain persistent identifiers (PIDs) where ever
possible (ORCID for persons, ROR for organisations, Re3Data for repositories, DOI for data 
products or documents). They also describe the data collection itself, as well as the 
interoperable interfaces that are implemented to access the collection. 

\section{FAIR Data publication and citation}
\begin{figure}
    \centering
    \includegraphics[width=\linewidth]{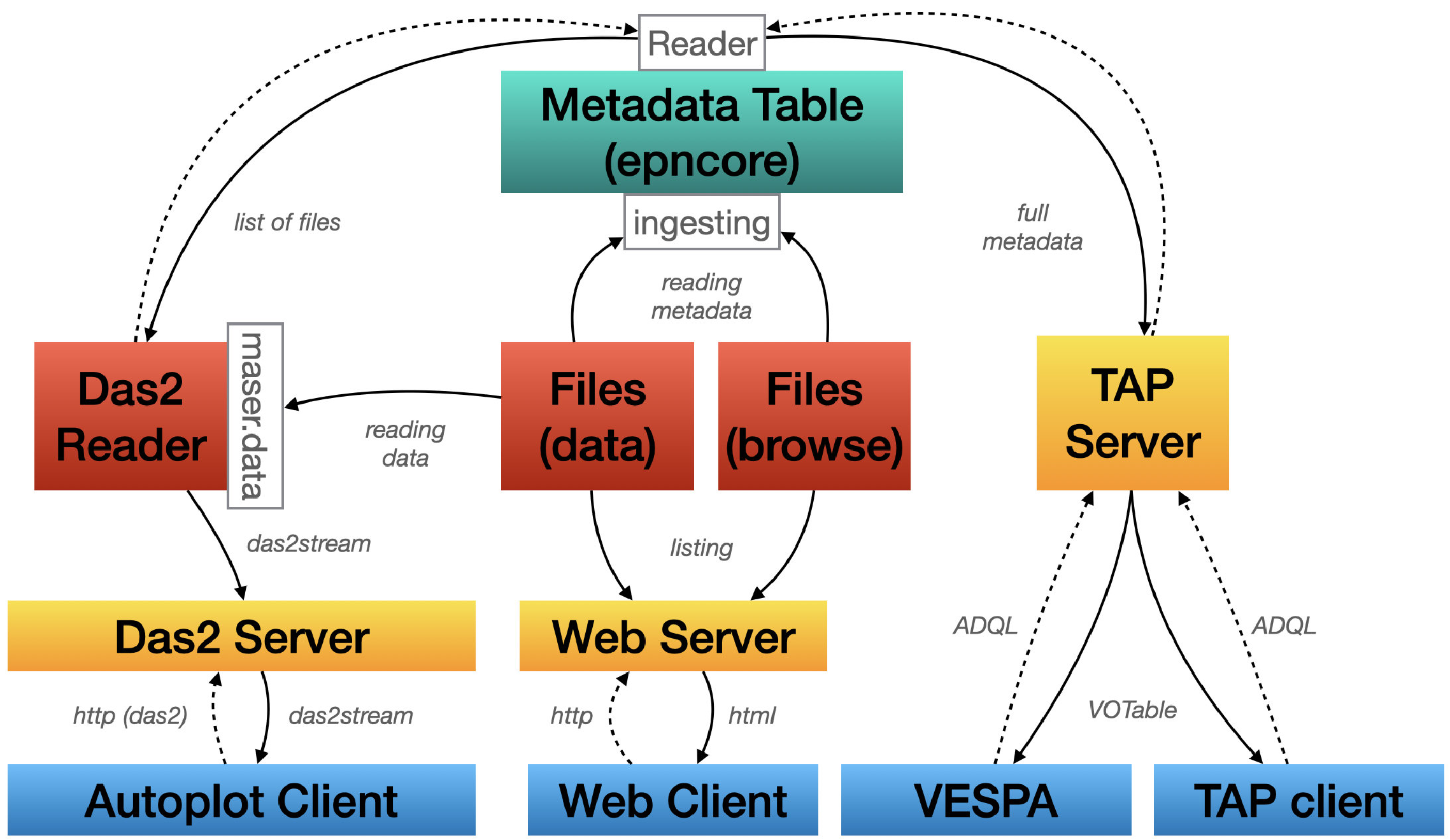}
    \caption{MASER data architecture. From right to left, the three access interfaces: data 
    discovery with \emph{EPN-TAP}, direct web access (including dataset landing page), and 
    remote data visualisation with \emph{Das2}. The \emph{epncore} table is shown at the top. 
    The \emph{das2} server-side scripts using \texttt{maser.data} are shown on the middle row, 
    as well as the data products and browse images stored on disk.}
    \label{fig:architecture}
\end{figure}

Figure \ref{fig:architecture} presents the overall picture of the data management 
architecture of the MASER service. It shows the various access interfaces at the bottom of 
the figure. At the top of the schema, the \emph{epncore} metadata table, which is used for 
several internal services, in addition to the VESPA interface. 

Using this set of tools, the MASER data products are thus \emph{Findable} (EPN-TAP data 
discovery interface; JSON-LD content using schema.org), \emph{Interoperable} (standard data 
format, such as FITS or CDF, and standard access interfaces), \emph{Accessible} (all products
are openly available, and are shared with a licence), and \emph{reusable} (documentation and 
metadata are associated to data products to help reuse, as allowed by their open licences).

The MASER landing pages have been assessed using FAIRness evaluation tools, such as 
FAIR-checker\footnote{\url{https://fair-checker.france-bioinformatique.fr}} 
\citep{rosnet_2022_5914307} and F-UJI\footnote{\url{https://www.f-uji.net}} 
\citep{anusuriya_devaraju_2020_4063720}. Each tool is evaluating the FAIR principles 
\citep{2016NatSD...360018W} in a different way, weighting the various items depending on the 
common usage of their community. FAIR-checker scores are rather high for MASER landing pages 
since we are using PIDs as much as possible. F-UJI scores are lower, because they expect usage 
of URI and terms from community ontologies, which are yet fully implemented in MASER related 
communities. 

The data citation policies varies across editors. Most publishers \citep{Chen_2022,stall_2023}
are now proposing to reference datasets and software in the references section of papers.   
The MASER service is proposing the tooling to enable data citation, and is encouraging authors 
to cite their data using the associated DOI. In the past years several papers have been 
published using MASER published data products (datasets, catalogues, supplementary materials). 
Three examples are given below, with data citations, as present in the published version.
\begin{itemize}
    \item \citet{Cecconi:2021tb} (published in Planetary and Space Sciences by Elsevier) 
    cites datasets from MASER and from NASA/PDS as well as supplementary
    material hosted by MASER. The reference section includes enough information (authors, year,
    title, version, publisher, DOI) for the data citation.     
    \item \citet{10.1029/2021ja030209} (published in Journal of Geophysical Research -- Space 
    Physics, by AGU) cites data products from MASER. The reference section is lacking key 
    information to properly identify data citation (only authors, year, title and version; no  
    publisher, no DOI). The missing information was sent to the editor in the \texttt{bibtex} 
    file but was not reported by the copy-editing staff, and the authors didn't pick this issue 
    during the proof-reading stage.
    \item \citet{10.1051/0004-6361/202140998} (published in Astronomy \& Astrophysics, by EDP 
    Sciences) cites a dataset. The reference section is lacking key 
    information to properly identify data citation (only authors, year and title; no  
    publisher, no DOI). The DOI is available in the ArXiV preprint (using the A\&A {\LaTeX} 
    template), but was not kept by the copy-editing staff.
    \item \emph{Note to the editors}, the default bibliographic {\LaTeX} template proposed for  
    this proceeding is a good example of what NOT to do for data and software referencing. No 
    DOI is associated to references, making it impossible to implement correct referecing to 
    data or software (such as the reference in this paper pointing to Zenodo).
\end{itemize}
At the time of writing of the paper, none of the metadata accessible through the 
\emph{Crossref} API are including DOI information on datasets for the three examples detailed 
previously. This is blocking citation tracking, or data publication impact evaluation. 

The NASA/ADS (Astrophysics Data System) bibliographic search engine team have implemented tools 
to overcome this missing information. For instance, the 
\emph{data availability statement} of the papers are parsed, and URLs are interpreted as data
references. However, when data is cited with a regular citation scheme in the reference section
of the paper, it is not (yet) included in the ADS page reference list (since it is not 
indexed in ADS), nor in the ADS links to data section. This results is an annoying situation
consisting in tracking URLs to data, which are perishable, and skipping data referenced with a 
persistent identifier. Direct submission from data providers using IVOA interfaces is also being 
tested to feed NASA/ADS with missing links to data. 

Data citation is part of scientific integrity as it facilitates science reproducibility. 
Scientists are not yet all implementing data citation in their work, despite the data citation
policies in place for most publishers. However, the short analysis presented in this paper 
shows that the data citation is currently not operating properly, even when the authors are 
making the efforts to provide the required information. Solutions exist, but these are just 
trying to fix the work not done by the editors.  

\section{Summary}

The MASER service team has selected a series of tools and community recommendations, which 
are implementing a FAIR ecosystem for time domain low frequency radioastronomy. The resulting 
toolbox is covering all the identified needs and is fully operational. Work has still to be 
done (including journal editors) to enable full data citation traceability, as show in the 
last section.

\section*{Note}
Some references in this manuscript \citep{annex_2018_2529131,
anusuriya_devaraju_2020_4063720,rosnet_2022_5914307,dmp_guide_obspm,dmp_template_obspm} are 
deliberately kept has produced by the editor's {\LaTeX} template (i.e., 
with insufficient reference information, such as publisher or DOI) to demonstrate the required 
editorial workflow changes. Adequate and findable (short) references are provided below:
\begin{itemize}
\item Annex, A., et al., 2018, v1.0, Zenodo, doi:10.5281/zenodo.2529131
\item Devaraju, A. \& Huber, R., 2020, v1.0.0, Zenodo, doi:10.5281/zenodo.4063720
\item Rosnet, T., et al., 2022, v1.0, Zenodo, doi:10.5281/zenodo.5914307
\item Stoll, V. et al., 2021, v1.0, Observatoire de Paris, doi:10.25935/1mh3-nn37
\item Stoll, V. et al., 2021, v1.0, Observatoire de Paris, doi:10.25935/x859-th79
\end{itemize}

\bibliography{I601} 

\end{document}